\documentclass[twocolumn,showpacs,amsmath,amssymb,prd]{revtex4}

\usepackage{graphicx}
\usepackage{epsfig}
\usepackage{dcolumn}
\usepackage{bm}
\usepackage{mathrsfs,amsmath}

\def\beq{\begin{equation}}
\def\enq{\end{equation}}
\def\ba{\begin{eqnarray}}
\def\ea{\end{eqnarray}}
\def\<{\langle}
\def\>{\rangle}

\urldef\lensnoise\url{http://lesgourg.web.cern.ch/lesgourg/codes.html}

\begin{document}

\title{Probing dark energy with redshift-drift}

\author{Matteo Martinelli$^{1,2}$, Stefania Pandolfi$^{3}$, C. J. A. P. Martins$^{4}$, P. E. Vielzeuf$^{4,5,6}$}

\affiliation{$^1$SISSA, Via Bonomea 265, Trieste, 34136, Italy}
\affiliation{$^2$INFN, Sezione di Trieste, Via Valerio 2, 34127 Trieste, Italy}
\affiliation{$^3$Dark Cosmology Centre, Niels Bohr Istitute, University of Copnehagen, Juliane Maries Vej 30 , 2100 Copenhagen, Denmark}
\affiliation{$^4$Centro de Astrof\'{\i}sica da Universidade do Porto, Rua das Estrelas, 4150-762 Porto, Portugal}
\affiliation{$^5$Faculdade de Ci\^encias, Universidade do Porto, Rua do Campo Alegre 687, 4169-007 Porto, Portugal}
\affiliation{$^6$Universit\'e Paul Sabatier---Toulouse III, 118 route de Narbonne 31062 Toulouse Cedex 9, France}

\begin{abstract}
Future redshift-drift measurements (also known as Sandage-Loeb signal) will be crucial to probe the so called ``redshift desert'', thus providing a new tool for cosmological studies. In this paper we quantify the ability of a future measurement of the Sandage-Loeb signal by a CODEX-like spectrograph to constrain a phenomenological parametrization of dynamical dark energy, specifically by obtaining constraints on $w_0$ and $w_a$. We also demonstrate that if used alongside CMB data, the Sandage-Loeb measurements will be able to break degeneracies between expansion parameters, thus greatly improving cosmological constraints.
\end{abstract}


\date{\today}
\maketitle

\section{Introduction}
The currently preferred model for our Universe, the $\Lambda$ Cold Dark Matter ($\Lambda$CDM) model, is in very good agreement with a range of observational probes, such as the Cosmic Microwave Background (CMB) anisotropies, the Large Scale Structure (LSS), the scale of the Baryonic Acoustic Oscillation (BAO) in the matter power spectrum and the luminosity distance of the Supernovae Type Ia (SNIa).

In particular, the latter probe has identified the accelerated expansion of the Universe \cite{Riess:1998cb,Perlmutter:1998np}. This apparently counter-intuitive behavior of the Universe is ascribed to a currently unknown component, dubbed Dark Energy (DE), which is dominant over the matter content and exhibits a negative pressure.  
Very little is known about this negative pressure component, but the simplest candidate for it is the Cosmological Constant: a constant energy density with a negative equation of state parameter (EoS, i.e. the ratio between the pressure of a component and its energy density), $w=P/\rho=-1$.

Although it is in accordance with all the existing cosmological observables, it has known theoretical problems, such as the \emph{``Why Now?''} and the \emph{``Fine Tuning''} problem. For these reasons, many other theoretical explanations for the DE have been proposed in which the EoS parameter evolves with time, although there is currently no firm observational evidence for such time evolution.
 
Lacking of an a-priori knowledge of the physical nature of the DE, a phenomenological parameterization in which the EoS evolves with time, such as the Chevallier-Polarski-Linder (CPL) \cite{Chevallier:2000qy,Linder:2002et} parameterization, is often used to get model independent constraints on the (possible) evolution of the EoS parameter. In this case, the EoS parameter can be written as
\begin{equation} 
w(z)=w_0+w_a \frac{z}{1+z}\,.
\end{equation}
If the EoS parameter varies with time, there should be a detectable departure from the evolution of the expansion rate $H(z)$ relative to the simpler $\Lambda$CDM model, in which the EoS parameter is constant and equal to $-1$ for all the times. Indeed, in the latter case the expansion rate of the universe $H(z)$, which evolves according to the Friedman equations,  can be written as:
\begin{equation}\label{HzLCDM}
\frac{H^2(z)}{H^2_0}=\Omega_m(1+z)^3+\Omega_{\Lambda}\,.
\end{equation}
where $\Omega_m$ and $\Omega_\Lambda$ are the matter and the Cosmological Constant energy density.
In the CPL parametrization it becomes
\begin{equation}
\frac{H^2(z)}{H^2_0}=\Omega_m\left(1+z\right)^3+\Omega_{DE}\left(1+z\right)^{3(1+w_0+w_a)}e^{-3\frac{w_az}{1+z}}\,.
\end{equation}

If one wants to constrain a possibly evolving dark energy EoS without \emph{a priori} assumptions on an underlying model, it is crucial to measure the expansion rate at many different redshifts. Among the known probes of the expansion rate, the CMB probes the expansion rate at redshift $z\sim1100$, while for much lower redshifts we could rely on Weak Lensing and BAO probes, and most noticeably on the  SNIa luminosity distance measurements. A potentially powerful new way of constraining the expansion history in the low-intermediate redshifts regime, up to $z\sim3$ could come in the next decade from the Active Galactic Nuclei (AGN) distance measurements \cite{Watson:2011um}, athough their reliability remains to be demonstrated.
 
In this paper we want to explore the constraining power on the Dark Energy EoS of a next-generation probe of the expansion history at intermediate redshifts, based on redshift-drift measurements \cite{Sandage,Loeb:1998bu}. In this paper we will refer to this probe as {\it Sandage-Loeb} test (SL) as first proposed by the authors of Ref. \cite{Corasaniti:2007bg}. The SL is a direct measurement of the temporal drift of distant sources redshift ($2\leq z \leq5$) due to the expansion of the Universe and, as first pointed out in \cite{Pasquini,Corasaniti:2007bg}, it has some important cosmological advantages as it is a direct probe of the evolution of the Universe, using relatively simple and well-understood physics to probe an otherwise inaccessible redshift range without any underlying theoretical assumptions (other than large-scale homogeneity and isotropy). A useful overview of SL properties can be found also in \cite{Vielzeuf:2012zd}.

The paper is organized as follows: in Section \ref{SL} we will revisit the theory behind the SL test for the considered models; in Section \ref{cod} we will summarize the observational requirements for the SL test; in Section \ref{am} we will describe the analysis method and the datasets we use in the present work; in Section \ref{res} we show our results and finally in Section \ref{concl} we derive our conclusions. 

\section{Sandage-Loeb Test}\label{SL}

The possibility of directly measuring the variation of the redshift of distant sources due to the expansion of the Universe was firstly conceived by Allan Sandage in 1962 \cite{Sandage}, but observational capabilities at that time would have failed to detect the small signal with a time interval between observations smaller than 10$^7$ years. Then in 1998, Abraham Loeb revisited the idea, and the conclusions of his studies were that, given the technological progress, it would have been possible to detect a drift of the redshift in the spectra of the Ly$\alpha$ forest of distant quasars (QSO) in the following few decades \cite{Loeb:1998bu}. In what follows we will briefly revisit the theory behind the Sandage-Loeb test, following the treatment of Ref. \cite{Corasaniti:2007bg}, and we propose some representative examples.

In a Friedmann-Robertson-Walker expanding Universe, the radiation emitted by a source at time $t_s$ and observed at time $t_0$ undergoes a redshift $z_s$ which is connected to the expansion rate through the scale factor $a(t)$ as
\begin{equation}
 1+z_s(t_0)=\frac{a(t_0)}{a(t_s)}.
\end{equation}
If we now consider the radiation emitted by the same source at the time $t_s+\Delta t_s$ and observed at $t_0+\Delta t_0$, the new redshift will be
\begin{equation}
 1+z_s(t_0+\Delta t_0)=\frac{a(t_0+\Delta t_0)}{a(t_s+\Delta t_s)}.
\end{equation}
With an adequate time interval between observations we can measure the difference in the observed redshifts $\Delta z_s$. This difference can be expressed, when $\Delta t<<t$, as \cite{Corasaniti:2007bg}
\begin{equation}
 \Delta z_s\approx\frac{\dot{a}(t_0)-\dot{a}(t_s)}{a(t_s)}\Delta t_0,
\end{equation}
but it is more conveniently expressed as a spectroscopic velocity shift through the relation
\begin{equation}
 \Delta v=\frac{c\Delta z_s}{1+z_s}
\end{equation}
where $c$ is the speed of light.\\
Therefore, using the first Friedmann equation and setting $a(t_0)=1$,  $\Delta v$ can be related to the matter content of the Universe \cite{Corasaniti:2007bg}
\begin{equation}
 \frac{\Delta v}{c}=H_0\Delta t\left[1-\frac{E(z_s)}{1+z_s}\right],
\end{equation}
where $H_0$ is the Hubble constant and $E(z)=H(z)/H_0$.

For the case of a flat $\Lambda$CDM Universe, from Eq. (\ref{HzLCDM}), $E(z_s)$ is
\begin{equation}\label{eq:estand}
 E(z_s)=\sqrt{\Omega_m\left(1+z_s\right)^3+\Omega_\Lambda}.
\end{equation}
Simple algebra shows that the velocity differences are very small, typically corresponding to a few $cm/s$, even assuming a time interval $\Delta t_0=30$ years between the observations (see for example Fig.\ref{fig:lambda}).

Different dark energy models will clearly yield different $\Delta v$ signals \cite{Balbi:2007fx}. In this paper we use the Chevallier-Polarski-Linder (CPL) parametrization \cite{Chevallier:2000qy,Linder:2002et} as our toy model. As described above the time dependence of $w$ changes the evolution of the Dark Energy energy density with respect to the cosmological constant case
\begin{equation}
 \Omega_\Lambda\Rightarrow\Omega_{DE}(z)=\Omega_{DE}^0e^{3\int_0^z{dz'\frac{1+w(z')}{1+z'}}},
\end{equation}
where $\Omega_{DE}^0=\Omega_{DE}(z=0)$; therefore, $E(z)$ becomes
\begin{equation}\label{HzCPL}
E(z)=\left[\Omega_m\left(1+z\right)^3+\Omega_{DE}\left(1+z\right)^{3(1+w_0+w_a)}e^{-3\frac{w_az}{1+z}}\right]^{1/2}\,.
\end{equation}

\section{CODEX and the Ly$\alpha$ forest of QSO}\label{cod}

The European Extremely Large Telescope (E-ELT) equipped with a high-resolution, ultra-stable spectrograph such as the COsmic Dynamics Experiment (CODEX) \cite{codex} will have the ability to detect the cosmological redshift drift in the Lyman $\alpha$ absorption lines of distant (2 $<z<$ 5) QSOs. In this Section we will briefly describe the underlying observational requirements.

As pointed out by the authors of \cite{Liske:2008ph} the accuracy in the determination of the tiny variation of the redshifts of the sources depends mostly on the signal-to-noise (S/N) ratio at which spectral features are detected, and in a photon-noise-limited experiment it depends, in turn, on (i) the size of the telescope; (ii) the flux density of the sources; (iii) the combined telescope/instrument efficiency and integration time. 

The new generation of $30+$m ELTs will be able to collect the large number of photons needed to efficiently resolve these features in the spectra of the selected sources in scientifically worthy amount of integration time. Concerning the second issue, Loeb suggested \cite{Loeb:1998bu} as a promising target the Lyman $\alpha$ forest of distant QSOs and more recently the authors of \cite{Liske:2008ph} quantitatively determined the accuracy of the detection of the cosmological redshift drift in the high-redshift quasars absorption lines, including  metal ones, in the era of the ELTs. They find that a 42-m ELT could achieve the accuracy needed to detect the redshift variation with a 4000 hours of integration in a period of $\Delta t_0=20$ years \cite{Liske:2008ph}.

The most suitable source targets for the SL test have been identified with the Ly$\alpha$ forest of the distant QSOs. The analysis of \cite{Liske:2008ph} has identified the following requirements for the target sources:
\begin{enumerate}
\item They should trace the Hubble flow, and have small peculiar motions.
\item They should have sharp spectral features.
\item They should have a large number of usable features to maximize the amount of information per unit observing time.
\item They should be as bright as possible.
\item They should exist over a wide redshift range, but in particular, at high redshifts, where the signal is expected to be higher.
\end{enumerate}
The Ly$\alpha$ forest lines meet all the above requirements, but the second one, because the intergalactic absorbing gas between us and the distant QSO is at a temperature of the order of 10$^4$ K (see e.g  \cite{Theuns:1999mz})  so that the absorption lines are not particularly sharp. Therefore we can use these astrophysical sources to detect the Sandage-Loeb signal, but in order to detect tiny shifts of the spectral lines, high-resolution, extremely stable (on timescales of tens of years) spectrographs are required.

\begin{figure}[h!]
\begin{center}
\hspace*{-1cm}
\begin{tabular}{cc}
\includegraphics[width=8cm]{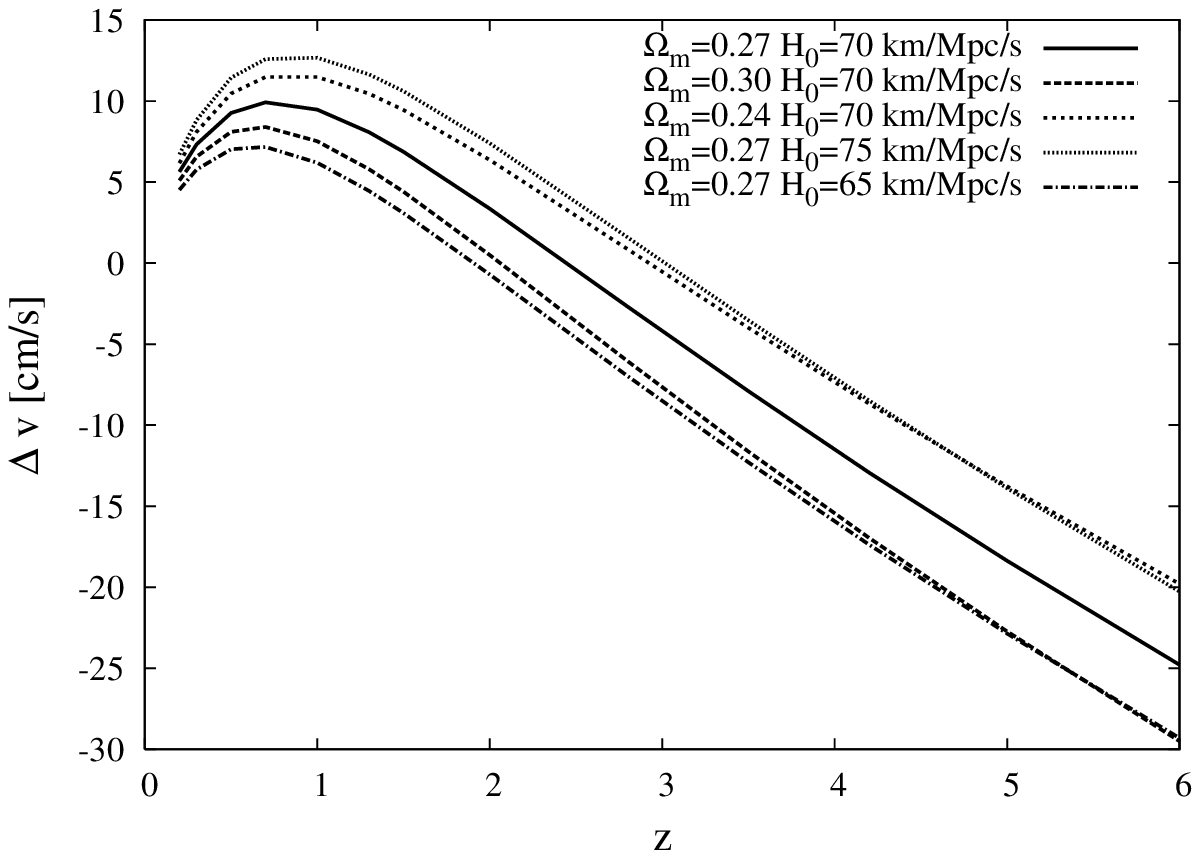}\\
\includegraphics[width=8cm]{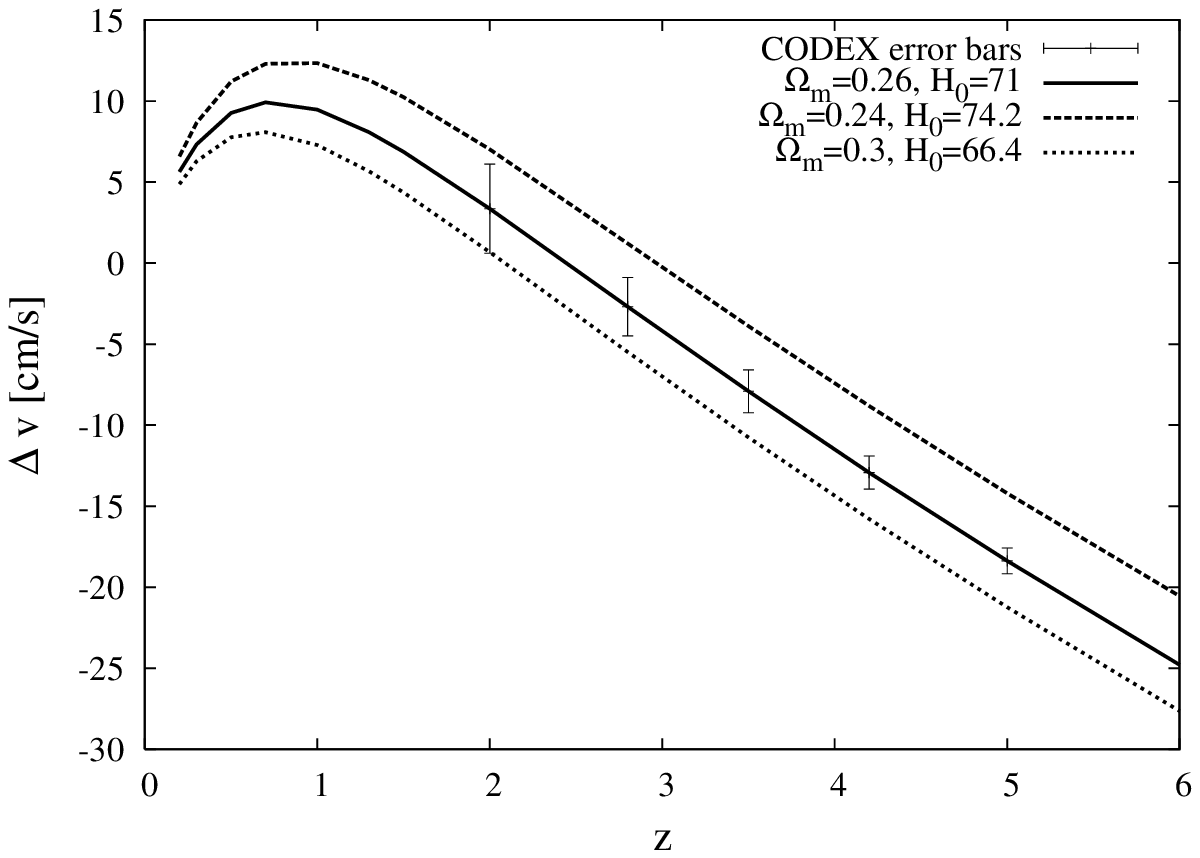}
 \end{tabular}
\caption{Top panel: $\Delta v$ signal for different values of $\Omega_m$ and $H_0$. Bottom panel: $\Delta v$ signal for different combinations of $\Omega_m$ and $H_0$ with fixed $\Omega_m\cdot H_0^2$ value. The error bars refers to the CODEX spectrograph.}
\label{fig:lambda}
\end{center}
\end{figure}

According to Monte Carlo simulations of Ref. \cite{codex}, CODEX has an error on the measured spectroscopic velocity shift $\Delta v$ that can be expressed as:
\begin{equation}\label{eq:error}
 \sigma_{\Delta v}=1.35\ \frac{2370}{S/N}\ \sqrt{\frac{30}{N_{QSO}}}\ \left(\frac{5}{1+z_{QSO}}\right)^x\ cm\ s^{-1},
\end{equation}
where $S/N$ is the signal to noise ratio, $N_{QSO}$ the number of observed quasars, $z_{QSO}$ their redshift and the exponent $x$ is equal to $1.7$ when $z\leq4$, while it becomes $0.9$ beyond that redshift. \\
In Fig.\ref{fig:lambda} we can notice how using these specifications it is already possible to distinguish between different cosmological models.

\begin{figure}[h!]
\begin{center}
\hspace*{-1cm}
\begin{tabular}{cc}
\includegraphics[width=8cm]{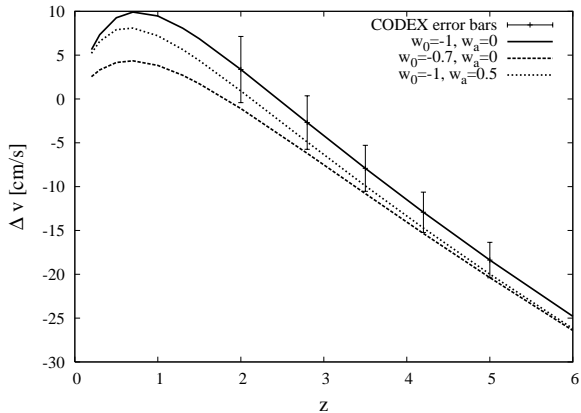}&
 \end{tabular}
\caption{$\Delta v$ signal for different combinations of EoS parameters $w_0$ and $w_a$. The error bars refers to the CODEX spectrograph.}
\label{fig:cpl}
\end{center}
\end{figure}

An important pro of the Sandage-Loeb test performed through the QSOs absorption lines is the redshift range of observations: suitable QSOs can be found at $2<z<5$ and this means that this test will probe a redshift region where there are no other cosmological probes \cite{Corasaniti:2007bg}, thus it can bring helpful informations to break degeneracies between cosmological parameters. It would be also very useful to detect the Sandage-Loeb signal at redshifts lower than $2$ as this would probe the Dark Energy dominated epoch in a complementart way with respect to other observables, e.g. SNIa. This is feasible in principle but such measurements have to be done in space and at present there is no plan for a space-based spectrograph with the needed sensitivity and stability. Nevertheless this probe allows in principle to distinguish between different Dark Energy models as can be seen in Fig.\ref{fig:cpl}.
Moreover SL test will be able to break degeneracies between cosmological parameter; as an example, observations of the CMB are sensitive to the product $\Omega_m \cdot H_0^2$, therefore combinations of this parameters with the same product value would be indistinguishable, while SL will be able to break this degeneracy as it probes the two parameters separately (see the bottom panel of Fig.\ref{fig:lambda}).

\section{Analysis Method}\label{am}
In this section we review the data simulation and the method used to obtain forecasted constraints on cosmological parameters. 
We sample the following  set of cosmological parameters, adopting flat priors on them: the baryon and cold dark matter densities $\Omega_{b}h^2$ and $\Omega_{c}h^2$, $\Theta_s$ that is the ratio between the sound horizon and the angular diameter distance at decoupling, and the two CPL parameters $w_0$ and $w_a$. We choose to fix non-geometrical cosmological parameters, $\tau$, $n_s$ and $A_s$, to their best-fit values as the Sandage-Loeb test is not sensitive to them.\\
We use the publicly available MCMC code \texttt{cosmomc} \cite{Lewis:2002ah} with a convergence
diagnostic using the Gelman and Rubin statistics. In principle we could have performed a simpler Fisher Matrix analysis \cite{Bassett:2009uv}, but we choose to perform a full MCMC analysis as it accounts for possible non-gaussian posteriors on sampled parameters.\\
The CPL parameters are sampled in \texttt{cosmomc} through the Parameterized Post-Friedmannn add-on of the \texttt{CAMB} package, which allows to cross the phantom divide ($w=-1$) \cite{Fang:2008sn}. We will not address this problem here, but it must be noticed that while the phenomenological parametrization allows to test these models, a careful theoretical treatment is needed when considering specific phantom models.
We modified the \texttt{cosmomc} code including an additional module able to compute the Sandage-Loeb signal for any value of the cosmological parameters and also able to calculate the likelihood of the resulting cosmological model, given a mock Sandage-Loeb dataset where the error bars are computed using Eq.(\ref{eq:error}), with a S/N of 3000 and a number of QSO $N_{QSO}=30$ assumed to be uniformly distributed among the following redshift bins $z_{QSO}=[2.0,2.8,3.5,4.2,5.0]$. 
The fiducial values for cosmological parameters are taken to be the best fit ones from WMAP seven years analysis (WMAP7) \cite{wmap7}, $\Omega_{b}h^2=0.02258$, $\Omega_{c}h^2=0.1109$, $\theta_s=1.0388$, and $w_0=-1$, $w_a=0$ for the CPL parameters.\\


We also built a full CMB mock dataset (temperature, E and B polarization modes) with noise properties consistent with a Planck-like~\cite{:2006uk} experiment (see Tab.~\ref{tab:exp} for the used specifications).\\

\begin{table}[!htb]
\begin{center}
\begin{tabular}{ccc}
Channel & FWHM & $\Delta T/T$ \\
\hline
70 & 14' & 4.7\\
100 & 10' & 2.5\\
143 & 7.1'& 2.2\\
\hline
& $f_{sky}=0.85$  & \\
\hline\hline

\end{tabular}
\caption{Specifications for a Planck-like experiment. Channel frequency is given in GHz, FWHM (Full-Width at Half-Maximum) in arc-minutes, and the temperature sensitivity per pixel in $\mu K/K$. The polarization sensitivity is $\Delta E/E=\Delta B/B= \sqrt{2}\Delta T/T$.}
\label{tab:exp}
\end{center}
\end{table}

The detector noise considered for each channel is $w^{-1}=(\theta\sigma)^2$, where $\theta$ is the FWHM (Full-Width at
Half-Maximum) with the assumption of a Gaussian beam profile and $\sigma$ is the temperature sensitivity $\Delta T$. Therefore, the following noise spectrum is added to fiducial $C_\ell$:
\begin{equation}
N_\ell = w^{-1}\exp(\ell(\ell+1)/\ell_b^2) \, ,
\end{equation}
where $\ell_b$ is given by $\ell_b \equiv \sqrt{8\ln2}/\theta$.\\
On top of these two probes, we also make use of the Hubble Space Telescope (HST) prior, $H_0=74.2\pm3.6$, available in \texttt{cosmomc} package \cite{Riess:2009pu}.\\

After computing the mock datasets we constrain the parameters introduced above using the Sandage-Loeb test and CMB alone and then we combine the two experiments to obtain joint constraints. The results are presented in the following section.

\section{Results}\label{res}
As stated in the previous section, we built 2 different datasets for the fiducial set of cosmological parameters, CMB mock spectra and a Sandage-Loeb mock dataset.
We first perform an analysis where $w_a$ is fixed to zero, using separately CMB and Sandage-Loeb, then we probe the same parameters using both probes to study the joint constraining power. In all these cases we add the HST prior.\\
\begin{figure}[h!]
\begin{center}
\hspace*{-1cm}
\begin{tabular}{cc}
\includegraphics[width=8.8cm]{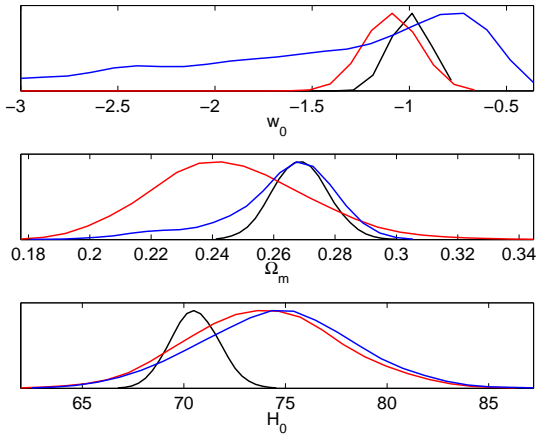}&
 \end{tabular}
\caption{1-D marginalized constraints on $\Omega_m$, $H_0$ and $w_0$ ($w_a$ fixed) considered using CMB (red lines), SL (blue lines) and CMB+SL (black lines).}
\label{fig:1d_wafixed}
\end{center}
\end{figure}

\begin{table}[!htb]
\begin{center}
\begin{tabular}{l|c|c|c}

               & CMB+HST & SL+HST & CMB+SL+HST \\
\hline
$\sigma{(\Omega_bh^2)}$       & $0.0001$   & $0.03$    & $0.0001$  \\
$\sigma{(\Omega_ch^2)}$       & $0.0006$   & $0.03$    & $0.0006$  \\
$\sigma{(\theta_s)}$          & $0.0003$   & $0.08$    & $0.0003$  \\
$\sigma{(w_0)}$               & $0.1$      & $unconstrained$       & $0.03$    \\
$\sigma{(H_0)}$               & $3.6$      & $3.6$     & $1.2$     \\
$\sigma{(\Omega_m)}$          & $0.02$     & $0.01$    & $0.009$   \\
\hline
\end{tabular}
\caption{$68 \%$ c.l. errors on cosmological parameters. $w_a$ is fixed to zero.}
\label{tab:results_wafixed}
\end{center}
\end{table}

\begin{figure}[h!]
\begin{center}
\hspace*{-1cm}
\begin{tabular}{cc}
\includegraphics[width=8cm]{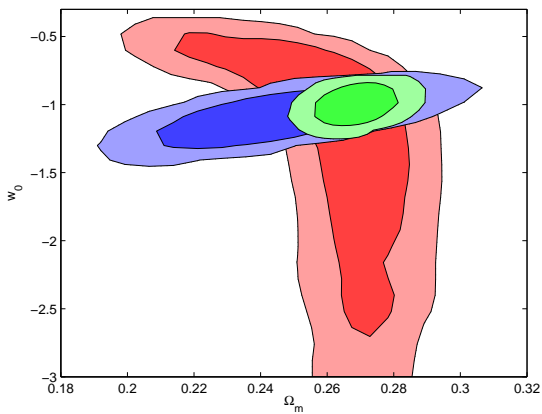}&
 \end{tabular}
\caption{2-D constraints on $w_0$ and $\Omega_m$ using CMB (blue), SL (red) and combining the two probes (green). }
\label{fig:2d_wafixed}
\end{center}
\end{figure}

\begin{figure}[h!]
\begin{center}
\hspace*{-1cm}
\begin{tabular}{cc}
\includegraphics[width=8cm]{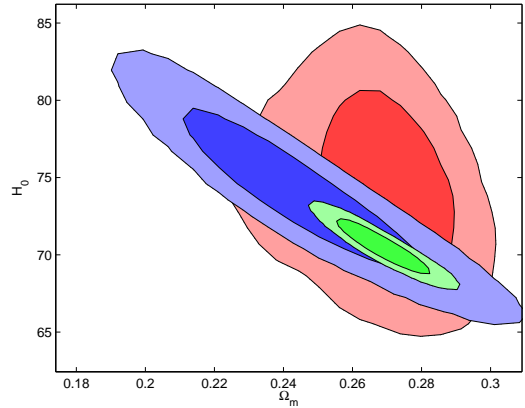}&
 \end{tabular}
\caption{2-D constraints on $H_0$ and $\Omega_m$ using CMB (blue), SL (red) and combining the two probes (green).}
\label{fig:2d_h0_wafixed}
\end{center}
\end{figure}

Table \ref{tab:results_wafixed} and Fig.~\ref{fig:1d_wafixed} show how combining these two probes greatly improves constraints on the geometrical cosmological parameters, even tough SL+HST alone are not able to constrain $w_0$, as very negative values of this parameter change the SL signal in a redshift range where CODEX is not sensitive.
We can see from  Figs.~\ref{fig:2d_wafixed}  and~\ref{fig:2d_h0_wafixed}  that the improvement  is not due to a better sensitivity of the SL test (SL constraints on $w_0$, $\Omega_m$ and $H_0$ are comparable to the ones obtained with CMB), but rather to different degeneracies; in these plot it is clear how the parameters $\Omega_m$, $w_0$ and $H_0$ present different directions of degeneracy. This implies that using SL together with CMB will break the degeneracies between the parameters and will improve the constraints.\\

\begin{figure}[h!]
\begin{center}
\hspace*{-1cm}
\begin{tabular}{cc}
\includegraphics[width=8cm]{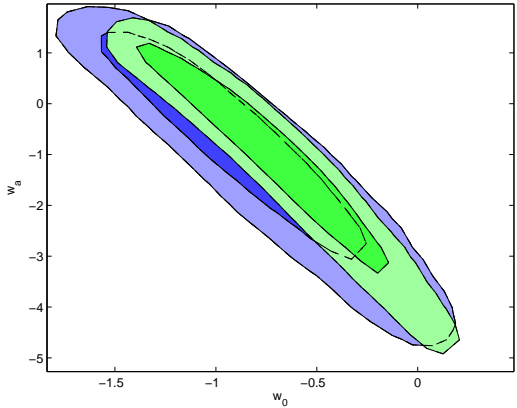}&
 \end{tabular}
\caption{2-D constraints on $w_0$ and $w_a$ using CMB alone (blue) and SL+CMB (green).}
\label{fig:2d_vary}
\end{center}
\end{figure}

We repeat the analysis with a $w_a$ free to vary, thus considering dynamical dark energy models. In this case we can notice from Table \ref{tab:results_wavary}  how the constraints on expansion parameters are not improved using SL in conjunction CMB, while on $H_0$ and $\Omega_m$ we have improvements in the constraining power as in the previous case. This means that the Sandage-Loeb test is not able to break the existing degeneracy between $w_0$ and $w_a$, as we can also see in Fig.\ref{fig:2d_vary}.

\begin{table}[!htb]
\begin{center}
\begin{tabular}{l|c|c}

               & CMB+HST & CMB+SL+HST \\
\hline

$\sigma{(\Omega_bh^2)}$       & $0.0001$   & $0.0001$  \\
$\sigma{(\Omega_ch^2)}$       & $0.0006$   & $0.0006$  \\
$\sigma{(\theta_s)}$          & $0.0003$   & $0.0003$  \\
$\sigma{(w_0)}$               & $0.4$      & $0.4$     \\
$\sigma{(w_a)}$               & $1.5$      & $1.5$     \\
$\sigma{(H_0)}$               & $3.7$      & $0.9$     \\
$\sigma{(\Omega_m)}$          & $0.03$     & $0.007$   \\
\hline
\end{tabular}
\caption{$68 \%$ c.l. errors on cosmological parameters when $w_a$ is free to vary..}
\label{tab:results_wavary}
\end{center}
\end{table}

\begin{figure}[h!]
\begin{center}
\hspace*{-1cm}
\begin{tabular}{cc}
\includegraphics[width=8cm]{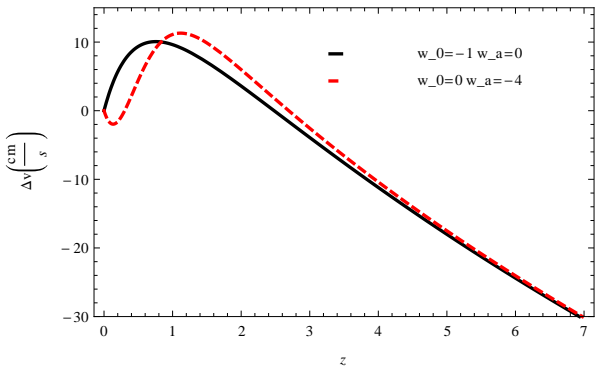}&
 \end{tabular}
\caption{SL signal for fiducial values of $w_0$ and $w_a$ and for values included in the $2-\sigma$ region of Fig.\ref{fig:2d_vary}.}
\label{fig:deltavwa}
\end{center}
\end{figure}

This result may seem somewhat puzzling as in Eq.(\ref{HzCPL}) there are two terms containing the CPL parameters: in the first term, the exponent $3(1+w_0+w_a)$, $w_0$ and $w_a$ are completely degenerate as they appear in a linear combination, while in the second one, only $w_a$ appear in an exponential term ($\exp{-3w_az_s/1+z_s}$).  Therefore, one may think that the latter term could be able to break the degeneracy. However it can be shown that this term only excludes very high values of $w_a$; in Fig.\ref{fig:deltavwa} we can see how using a set ($w_0=0,\ w_a=-4$), allowed by CMB+SL, we obtain an SL signal that does not significantly differ from the fiducial one, compared to the error bars of Fig.\ref{fig:lambda}. An even higher sensitivity of future experiments to the SL signal in the moderate redshift range would be needed to break this degeneracy. However, an alternative way to do this would be to probe also the low-redshift range with SL measurements but, as previously pointed out would require measurements from space.

\section{Conclusions}\label{concl}

In this paper we evaluated the constraining power of the Sandage-Loeb test on the phenomenological dynamic Dark Energy parameters $w_0$ and $w_a$ and on other expansion-related cosmological parameters, such as $\Omega_m$ and $H_0$. We explored both the possibility of a constant dark energy EoS ($w_a$ fixed to zero) and a more general case where $w_a$ is free to vary.

In the first case we found 1-$\sigma$ errors that are competitive with the ones obtained using forecasted data for a Planck-like CMB experiment. Moreover, we highlighted how Sandage-Loeb observations alongside CMB data can break degeneracies between different parameters; we obtained that the use of Sandage-Loeb probe can improve CMB results on $w_0$, $H_0$ and $\Omega_m$ by factors of  $3.3$, $3$ and $2.2$ respectively. However, in the second case, where $w_a$ is a free parameter, we found that Sandage-Loeb is not able to remove the degeneracy between $w_0$ and $w_a$ and therefore it cannot improve CMB constraints on these parameters; nevertheless, significantly tighter constraints on $H_0$ and $\Omega_m$ are again achieved.

However we should point out that these results are obtained assuming the absence of systematic effects, which play a key role in the measurement of Dark Energy parameters. Therefore a precise understanding of these will be necessary in order to obtain reliable predictions on future constraints. For some caveats on the effect of systematics on cosmological constraints see\cite{Linder:2008pp,Linder:2010qn}.

Nevertheless, we expect that a future detailed analysis of specific dark energy models should be able to show how Sandage-Loeb can be crucial in eventually ruling out these models, as many of them have specific characteristics leading to a discernible $\Delta v$ signal (see e.g. \cite{Quercellini:2010zr,Vielzeuf:2012zd}) in the redshift range of interest.

\section{Aknowledgments}

This work was done in the context of the joint Master in Astronomy of the Universities of Porto and Toulouse, supported by project AI/F-11 under the CRUP/Portugal--CUP/France cooperation agreement (F-FP02/11). CJM and PEV acknowledge the support of project PTDC/FIS/111725/2009 from FCT (Portugal), and additional support from grant PP-IJUP2011-212 (funded by U. Porto and Santander-Totta). The Dark Cosmology Centre is funded by the Danish National Research Foundation. The work of CJM is supported by a Ci\^encia2007 Research Contract, funded by FCT/MCTES (Portugal) and POPH/FSE (EC). We thank Pier Stefano Corasaniti and Eric Linder for useful corrections and comments.


\begin{thebibliography}{srt}

\bibitem{Riess:1998cb} 
  A.~G.~Riess {\it et al.}  [Supernova Search Team Collaboration],
  Astron.\ J.\  {\bf 116}, 1009 (1998)
  [astro-ph/9805201].
\bibitem{Perlmutter:1998np} 
  S.~Perlmutter {\it et al.}  [Supernova Cosmology Project Collaboration],
  Astrophys.\ J.\  {\bf 517}, 565 (1999)
  [astro-ph/9812133].

\bibitem{Chevallier:2000qy}
  M.~Chevallier and D.~Polarski,
  Int.\ J.\ Mod.\ Phys.\ D {\bf 10} (2001) 213
  [gr-qc/0009008].
\bibitem{Linder:2002et}
  E.~V.~Linder,
  Phys.\ Rev.\ Lett.\  {\bf 90} (2003) 091301
  [astro-ph/0208512].


\bibitem{Watson:2011um} 
  D.~Watson, K.~D.~Denney, M.~Vestergaard and T.~M.~Davis,
  Astrophys.\ J.\  {\bf 740}, L49 (2011)
  [arXiv:1109.4632 [astro-ph.CO]].
\bibitem{Sandage}
A.~Sandage,
''The Change of Redshift and Apparent Luminosity of Galaxies due to the Deceleration of Selected Expanding Universes.''
Ap.J. 136, 319-33 (1962).

\bibitem{Loeb:1998bu}
  A.~Loeb,
  Astrophys.\ J.\  {\bf 499} (1998) L111
  [astro-ph/9802122].

\bibitem{Corasaniti:2007bg}
  P.~S.~Corasaniti, D.~Huterer and A.~Melchiorri,
  Phys.\ Rev.\ D {\bf 75} (2007) 062001
  [astro-ph/0701433].

\bibitem{Pasquini}
Pasquini L., et~al., 2005, The Messenger, 122, 10

\bibitem{Vielzeuf:2012zd}
  P.~E.~Vielzeuf and C.~J.~A.~P.~Martins,
  Phys.\ Rev.\ D {\bf 85} (2012) 087301
  [arXiv:1202.4364 [astro-ph.CO]].






\bibitem{Theuns:1999mz} 
  T.~Theuns, J.~Schaye and M.~Haehnelt,
  Mon.\ Not.\ Roy.\ Astron.\ Soc.\  {\bf 315}, 600 (2000)
  [astro-ph/9908288].




\bibitem{codex}
CODEX Phase A Science Case, document E-TRE-IOA-573-0001 Issue 1 (2010)


\bibitem{Liske:2008ph} 
  J.~Liske, A.~Grazian, E.~Vanzella, M.~Dessauges, M.~Viel, L.~Pasquini, M.~Haehnelt and S.~Cristiani {\it et al.},
  Mon.\ Not.\ Roy.\ Astron.\ Soc.\  {\bf 386}, 1192 (2008)
  [arXiv:0802.1532 [astro-ph]].
  
  
  





\bibitem{Balbi:2007fx}
  A.~Balbi and C.~Quercellini,
  Mon.\ Not.\ Roy.\ Astron.\ Soc.\  {\bf 382}, 1623 (2007)
\bibitem{Quercellini:2010zr}
  C.~Quercellini, L.~Amendola, A.~Balbi, P.~Cabella and M.~Quartin,
  arXiv:1011.2646 [astro-ph.CO].

\bibitem{Lewis:2002ah}
A. Lewis and S. Bridle,
Phys.\ Rev.\ D {\bf 66}, 103511 (2002) (Available from
\texttt{http://cosmologist.info}.)
\bibitem{Bassett:2009uv}
  B.~A.~Bassett, Y.~Fantaye, R.~Hlozek and J.~Kotze,
  Int.\ J.\ Mod.\ Phys.\ D {\bf 20} (2011) 2559
  [arXiv:0906.0993 [astro-ph.CO]].

\bibitem{Fang:2008sn}
  W.~Fang, W.~Hu and A.~Lewis,
  Phys.\ Rev.\ D {\bf 78} (2008) 087303
  [arXiv:0808.3125 [astro-ph]].


\bibitem{wmap7}
 E.~Komatsu {\it et al.}  [WMAP Collaboration],
  Astrophys.\ J.\ Suppl.\  {\bf 192}, 18 (2011)
  [arXiv:1001.4538 [astro-ph.CO]]~;
\bibitem{:2006uk}
    [Planck Collaboration],
  arXiv:astro-ph/0604069.
  
\bibitem{Riess:2009pu}
  A.~G.~Riess {\it et al.},
  Astrophys.\ J.\  {\bf 699} (2009) 539
  [arXiv:0905.0695 [astro-ph.CO]].





\bibitem{Linder:2008pp}
  E.~V.~Linder,
  Rept.\ Prog.\ Phys.\  {\bf 71} (2008) 056901
  [arXiv:0801.2968 [astro-ph]].

\bibitem{Linder:2010qn}
  E.~V.~Linder,
  arXiv:1004.4646 [astro-ph.CO].


\end{thebibliography}
\end{document}